\documentclass[compsoc,conference,a4paper,10pt]{IEEEtran}
\IEEEoverridecommandlockouts
\usepackage{cite}
\usepackage{amsmath,amssymb,amsfonts}
\usepackage{graphicx}
\usepackage{textcomp}
\usepackage{bmpsize}
\usepackage{xcolor}
\usepackage{lipsum}
\usepackage[colorlinks=true,urlcolor=black]{hyperref}

\usepackage{svg}
\usepackage{listings}
\usepackage{mdframed}
\usepackage{mathtools}
\usepackage{amsfonts}
\usepackage[capitalize]{cleveref}
\usepackage{url}
\usepackage{algorithm}
\usepackage{algpseudocode}
\usepackage[nolist, nohyperlinks]{acronym}

\usepackage{csquotes}
\usepackage{xspace}
\usepackage{siunitx}

\newcommand{\bheading}[1]{{{\textbf{#1}}}}  

\renewcommand{\texttt}[1]{$\mathtt{#1}$}

\usepackage{color}
\definecolor{halfred}{HTML}{800000}
\definecolor{halfgreen}{HTML}{008000}
\definecolor{codeindent}{HTML}{cccccc}

\newcommand{\ecall}{\texttt{ecall}\xspace}
\newcommand{\ecalls}{\texttt{ecall}s\xspace}
\newcommand{\eapp}{\texttt{eapp}\xspace}
\newcommand{\eapps}{\texttt{eapps}\xspace}
\newcommand{\hpcfa}{HPCCFA\xspace}
\newcommand{\hellotracee}{\texttt{hello}-tracee\xspace}
\newcommand{\tweetnacltracee}{\texttt{tweetnacl}-tracee\xspace}
\newcommand{\dyndistracee}{\texttt{dd}-tracee\xspace}

\begin{acronym}
\acro{hpc}[HPC]{Hardware Performance Counter}
\acro{cfg}[CFG]{Control Flow Graph}
\acro{bb}[BB]{Basic Block}
\acro{cfa}[CFA]{Control Flow Attestation}
\acro{ilp}[ILP]{Integer Linear Programming}
\acro{tee}[TEE]{Trusted Execution Environment}
\acro{sm}[SM]{Security Monitor}
\acro{csr}[CSR]{Control and Status Register}
\acro{cfi}[CFI]{Control Flow Integrity}
\acro{tcb}[TCB]{Trusted Computing Base}
\acro{ipc}[IPC]{Inter Process Communication}
\acro{pmp}[PMP]{Physical Memory Protection}
\acro{rop}[ROP]{Return-Oriented Programming}
\acro{jop}[JOP]{Jump-Oriented Programming}
\acro{pebs}[PEBS]{Processor Event-Based Sampling}
\acro{cbcsolver}[CBC]{COIN-OR Branch and Cut solver}
\end{acronym}

\def\BibTeX{{\rm B\kern-.05em{\sc i\kern-.025em b}\kern-.08em
    T\kern-.1667em\lower.7ex\hbox{E}\kern-.125emX}}
\begin{document}

\title{HPCCFA: Leveraging Hardware Performance Counters for Control Flow Attestation}

\author{\IEEEauthorblockN{Claudius Pott} 
\IEEEauthorblockA{\textit{University of L\"ubeck} \\
L\"ubeck, Germany \\
c.pott@uni-luebeck.de}
\and
\IEEEauthorblockN{Luca Wilke}
\IEEEauthorblockA{\textit{Azure Research} \\
\textit{Microsoft} \\
Cambridge, UK\\
l.wilke@uni-luebeck.de}
\and
\IEEEauthorblockN{Jan Wichelmann}
\IEEEauthorblockA{\textit{University of L\"ubeck} \\
L\"ubeck, Germany \\
j.wichelmann@uni-luebeck.de}
\and
\IEEEauthorblockN{Thomas Eisenbarth}
\IEEEauthorblockA{\textit{University of L\"ubeck} \\
L\"ubeck, Germany \\
thomas.eisenbarth@uni-luebeck.de}
}
\maketitle

\begin{abstract}
\acp*{tee} allow the secure execution of code on remote systems without the need to trust their operators. They use static attestation as a central mechanism for establishing trust, allowing remote parties to verify that their code is executed unmodified in an isolated environment. 
However, this form of attestation does not cover runtime attacks, where an attacker
exploits vulnerabilities in the software inside the \ac{tee}.
\ac*{cfa}, a form of runtime attestation, is designed to detect such attacks.

In this work, we present a method to extend \acsp*{tee} with \acs*{cfa} and discuss how it can prevent exploitation in the event of detected control flow violations.
Furthermore, we introduce \hpcfa, a mechanism that uses \acp*{hpc} for \acs*{cfa} purposes, enabling hardware-backed trace generation on commodity CPUs.
We demonstrate the feasibility of \hpcfa on a proof-of-concept implementation for Keystone on RISC-V.
Our evaluation investigates the interplay of the number of measurement points and runtime protection, and reveals a trade-off between detection reliability and performance overhead.
\end{abstract}


\section{Introduction}
The use of \acfp{tee}, especially in cloud settings, is becoming ever more prevalent. \acp{tee} allow the execution of code in \emph{enclaves} and provide a range of security guarantees, such as isolation, confidentiality, and integrity. 
A central security mechanism is static attestation, where the \ac{tee} measures the state of the enclaves after initialization, thus being able to attest this initial state to remote parties. This prevents the host OS from tampering with the enclave's binary, as any modifications would show up in the attestation report. 

However, these mechanisms are not suitable to detect or prevent runtime attacks, where an enclave is targeted after its initialization. Prominent examples of runtime attacks are \ac{rop} and \ac{jop} attacks.
Runtime attestation provides the means to detect such attacks. One runtime attestation concept is \ac{cfa}, where the goal is to ensure that a program's execution follows the expected control flow.
A long line of work \cite{cflat2016,lofat2017,atrium2017,litehax2018,scarr2019,lahel2020,recfa2021} explores different approaches for \ac{cfa}. Most of them rely on a remote verifier \cite{cflat2016,lofat2017,atrium2017,litehax2018,scarr2019,recfa2021}, and a common assumption is a trusted operating system \cite{scarr2019,recfa2021}.
The approach proposed by \cite{guarantee2023} is related to the concept presented in this paper. It also uses two enclaves running on the same system to enable and protect its attestation.
However, their design relies on the components of the traced enclave for parts of the trace generation, which could be problematic if it is fully compromised. Furthermore, they are creating their \acp{cfg} by tracing benign executions of their program, which introduces the risk of falsely rejecting valid control flows, due to an incomplete \ac{cfg}.

In this work, we present \hpcfa, a \ac{cfa} design for \acp{tee}.
It allows live verification of code running in enclaves and is able to prevent leakages before they take effect. The design also incorporates a clean separation of the traced code and the generation of the necessary traces, preventing fully compromised enclaves from tampering with their traces. Additionally, the verification is performed on the same physical system, avoiding any latencies incurred by sending the traces through a network. Lastly, the design does not require any hardware modifications, making it applicable to generic and commonly available hardware.

The key to accomplish this is to run the attested code in one enclave and have a connected enclave responsible for the attestation.
Our design requires only minimal changes to the \ac{tee} to enable mutual authentication between two enclaves and provide trace generation capabilities, while integrating well with existing approaches to gather control flow data.
However, these approaches require either software instrumentation, which could potentially be overcome by an attacker, or hardware changes, which exclude existing hardware, to trace the control flow.

Therefore, as a second contribution, we present \hpcfa. It implements our \ac{cfa} design and uses the existing \acl{hpc} architecture to generate traces for the \acl{cfa}. It does not require hardware changes and does not debend on the monitored application to generate its own traces.
To this end, the \ac{tee} takes snapshots of the performance counter values every time control passes to or from the traced enclave. These snapshots are then passed to the tracer enclave which performs its verification. This relies on the \ac{cfg} of the traced enclave in combination with accurate \ac{hpc} predictions for each transition. The approach is designed to never reject valid control flows, thus eliminating the risk of false alarms and erroneous program terminations.
We provide a proof-of-concept implementation of \hpcfa for Keystone on RISC-V using the quad core StarFive VisionFive2 SoC.

Our evaluation shows that the concept works effectively for shorter or less complex program sequences, although it also shows a low detection rate for long and complex program sequences
Based on these insights, we explore how the addition of measurement points can keep all sequences of an enclave short and, therefore, allow reliable \ac{cfa} for entire enclaves.

\subsection{Our Contribution}
For our main contributions, we:
\begin{itemize}
    \item propose an architecture for flexible on device \ac{cfa} and prototype it in Keystone
    \item present \hpcfa, an \ac{hpc}-based \ac{cfa} scheme and analyze its efficacy and performance
    \item discuss use cases of \ac{hpc} for \ac{cfa}.
\end{itemize}
Our source code will be made available on Github. 

\noindent\bheading{Outline.}
This paper is structured as follows. We begin with a summary of relevant background information (\Cref{sec:background}). Then follow up with an introduction to our general system design (\Cref{sec:system-design}) and the more concrete variant using \acp{hpc} (\Cref{sec:hpc_for_cfa}) that was implemented (\Cref{sec:implementation-in-keystone}) and evaluated (\Cref{sec:evaluation}) in this paper. Afterwards, we analyze the security (\Cref{sec:security-analysis}) and discuss the approach and future work (\Cref{sec:discussion}). A discussion of related work (\Cref{sec:relaated-work}) will follow before closing with a conclusion (\Cref{sec:conclusion}).

\section{Background}\label{sec:background}
We will begin with an introduction to runtime attestation, followed by  information on \acp{tee} and specifically on Keystone. We will then close this section with an introduction into \acp{hpc}.

\subsection{Runtime Attestation}
As previously noted, one form of runtime attestation is \ac{cfa}. Here, the system is usually split into two parties. One is the application whose control flow we want to attest, the other is an application that given some kind of control flow trace has to decide whether the trace complies with the expected control flow. Naturally, the former party is viewed as untrusted, the latter is viewed as trusted.

The concepts of \ac{cfa} and \ac{cfi} are closely related.
The difference between the two is that \ac{cfa} is providing a way to verify that the expected control flow was actually followed, while \ac{cfi} mechanisms are preventing certain control flow attacks with additional security layers, e.g., Intel CET\footnote{\url{https://www.intel.com/content/www/us/en/developer/articles/technical/technical-look-control-flow-enforcement-technology.html} (Accessed 2025-04-04)}. 
We consider \hpcfa to be a \ac{cfa} approach, because it does not prevent control flow violations. However, the aspect of detecting such violations paired with the ability to prevent a compromised enclave from leaking any data also shows parallels to \ac{cfi} approaches.

Other mechanisms in the field of runtime attestation are not based on the control flow of the attested programs. Examples for these are \cite{smile2022} and \cite{triglav2021}. While \cite{smile2022} introduces a concept that allows memory introspection for SGX enclaves without breaking their security guarantees, \cite{triglav2021} utilizes SGX to implement a component running on the hypervisor that can attest the runtime integrity of VMs to remote users.

\subsection{Trusted Execution Environments}
The concept of \acp{tee} has been around for more than two decades. They provide CPUs with the capabilities to create an isolated execution environment that can be used to securely execute code, without the need to trust the host OS. 

Typical \ac{tee} applications have two separate components - the host application and the enclave. The host application runs within the context of the host OS, as any ordinary application would. Thus, it does not provide strong isolation guarantees, instead it is used to setup the enclave and serves as a communication interface between the trusted and untrusted world.
In contrast, the enclave runs inside the isolated execution environment, providing strong protections even against the host OS. However, it also means that it is not able to access basic system resources, such as network interfaces or the host OS's file system.
When developing a trusted application, one defines an edge call (\ecall) interface between the host application and its enclave. This interface can provide the enclave with access to system resources and allows the host application to send input to the enclave or receive output from it. Usually, this form of communication is enabled by the use of shared memory.
Some tasks have to be completed by the \ac{tee} itself, such as cleaning up registers when switching out of a trusted context or blocking access to certain memory regions. This usually involves a trusted software component that we refer to as the \acf{sm}.

\subsection{Keystone}\label{sec:background-keystone}
Keystone~\cite{DBLP:conf/eurosys/LeeKSAS20} is an open-source \ac{tee} for RISC-V.
It uses a programming model similar to Intel SGX, where the enclave is split into two parts -- the \textit{host app} and the \textit{eapp}. 
The main task of the host application is to facilitate the loading of the \eapp as well as to act as a proxy for the \eapp to interact with the host system. The \eapp can make use of both S-mode and U-mode, allowing it to set up its own runtime environment (e.g. Eyrie \cite{DBLP:conf/eurosys/LeeKSAS20} or SeL4\footnote{\url{https://sel4.systems/} (Accessed 2025-11-20)}).
The \textit{\acf{sm}} acts as the root of trust of the system by controlling the M-mode privilege level. It offers two APIs, one for the untrusted host apps and one that is available only to the trusted \eapps. The former is used, for example, to create a new \eapp while the latter can allow an \eapp to engage in a nonce based protocol to retrieve its attestation report.

To isolate \eapps from the remaining system, Keystone uses the optional \textit{\ac{pmp}}~\cite[Sec 3.7]{riscvPrivileged} feature of RISC-V. This feature allows M-mode software, such as Keystone's Security Monitor, to configure which physical address regions are accessible by the lower privilege levels. The \ac{sm} updates the \ac{pmp} rules when context switching into and out of \eapps to achieve its isolation guarantees.

\subsection{Hardware Performance Counters}
\acp{hpc} also referred to as Hardware Performance Monitoring Counters, are counting units built into modern CPUs. They allow tracking a wide variety of hardware events during the CPU's normal execution.
\acp{hpc} were originally designed to provide developers with information about potential bottlenecks in their code, but have since been used to accomplish variety of tasks, e.g. Malware Detection \cite{DBLP:journals/taco/WangCILK16,DBLP:conf/ccs/ZhouGJEJ18}.

In this work, we aim to accurately predict performance counter values for certain code blocks. We therefore differentiate between \emph{deterministic} and \emph{non-deterministic} performance counters, based on their behavior when running the same code multiple times. For example, information about cache hits and misses can differ between multiple runs of the same code, while the number of retired instructions and branches taken will always be the same.
In this work we focus only on deterministic counters, as this allows us to exactly predict the performance counter values purely based on the executed code.

\section{System Design} \label{sec:system-design}
In this section, we give an overview of the general system design of our \ac{cfa} approach.

As outlined in the introduction, we designed our system with the following goals:
\begin{enumerate}
    \item Live verification of control flow data with low runtime overhead,
    \item Clean separation of traced code and control flow trace generation,
    \item Verification of control flow traces on the same physical system that the traced code runs on,
    \item No hardware modifications.
\end{enumerate}

This section will first explain our threat model and then give an overview of the concept on how to extend enclaves with \ac{cfa} capabilities, followed by details on the trace generation and verification workflow.

\subsection{Threat Model}
We assume that the \ac{tee} provides the expected security guarantees and is implemented correctly, i.e. free from bugs. More concretely, we assume that the \ac{tee} performs static attestation on the initial contents of all enclaves to verify that they contain the expected code and that the attacker cannot directly read from or write to the memory of the enclave after initialization.
We also assume that the hardware implementation of the system is bug-free, including the implementation of components like the \ac{hpc} architecture and the \ac{pmp}.

The goal of our attacker is to leak secret data that is processed inside an enclave, so they have full control over the host operating system.
Additionally, we assume that the software inside the traced enclave has vulnerabilities that allow an attacker to arbitarily manipulate the control flow via an externally accessible API.

In this work, we are not considering side-channel attacks or hardware attacks to be in scope as these present an orthogonal issue. Further, we are assuming that the attacker has no interest in performing a Denial of Service attack, as this would be trivial for an OS-level attacker.

\subsection{General Runtime Attestation Concept}
Throughout this paper, we will call the application whose control flow we want to monitor the \textit{tracee}; its counterpart, which is verifying the validity of the control flow data, is called the \textit{tracer}.
The control flow data that needs to be verified is also referred to as (control flow) trace.
In our design, the tracer and the tracee are running inside their own enclaves inside a \ac{tee}. This has various advantages for both components.

Running the tracee in an enclave drastically improves the security guarantees that we can achieve with a control flow trace. While the static attestation mechanism of the \ac{tee} allows us to prove the integrity of the tracee's code at load time, the \ac{tee}'s isolation guarantees ensure that no external code can interfere with our application's execution. Thus, all of the code that can potentially interfere with the execution of the tracee is proven to be initially benign and is continuously monitored afterwards.

We need to evaluate the tracee's control flow traces in a trustworthy environment. Commonly, this is done by sending them to a remote system, where the evaluation can happen independently \cite{cflat2016,lofat2017,atrium2017,litehax2018,scarr2019,lahel2020,recfa2021}. When a (semi-)live verification of the control flow is desired, this adds a lot of latency. Since we already use a \ac{tee} for the tracee, we can avoid this bottleneck by running the tracer in an enclave as well. This fulfills the need for a trusted evaluation environment even on an untrusted system, thus enabling us to run both tracer and tracee on the same system, without compromising the security. This also allows us to use shared memory for low-latency and high-bandwidth transfer of the control flow traces.

Immediately attesting the validity of each control flow instruction would add an impractical amount of overhead. Instead, we suggest leveraging the isolation model of \acp{tee} to perform the verification in a slightly delayed, batched manner \emph{without} increasing the security implications of a successful control flow attack inside the tracee's enclave.
The key insight to achieve this is that any malicious behavior of the tracee, e.g. as a result of a breached control flow, is acceptable as long as we are able to detect it before it can change any state outside of the boundary of the tracee's enclave.
We capitalize on this insight by prohibiting access to any shared memory areas between the enclave and the remaining untrusted system, until the control flow status of the enclave has been successfully verified. This fits well into the already established \ecall programming model for secure enclaves.

It is vital that the tracer and tracee enclaves are always executed in tandem, while neither is oblivious to the other. 
The tracee's control flow information is security sensitive; therefore, it may only be read by its trusted tracer enclave. 
The tracee ensures that only a trusted tracer can be connected by specifying the tracer's hash in its own metadata, which is covered by the static attestation and can be verified externally.
On the other hand, a tracee enclave is only allowed to run while its tracer is attached, ensuring that the \ac{cfa} has to be enabled.

While it may be feasible to have one tracer that is able to attest multiple tracees even at the same time, doing so would not be practical, as this would require the tracer to be updated if any of its tracees is updated, which in turn would require all of the tracees to be updated with the tracers new hash.

\subsection{Trace Generation}\label{sec:trace-generation}
Our tracer/tracee system design is not bound to a specific mechanism to obtain control flow data.
On a high level, we simply require the mechanism to write the control flow data to a memory buffer managed by the \ac{sm} that is inaccessible to any lower privileged software, especially the tracee.
As detailed in the next section, the \ac{sm} will forward the data to the tracer enclave and orchestrate the attestation process.

In this paper, we explore the feasibility of using \acp{hpc} to gather control flow data.
To this end, the \ac{sm} manages the \acp{hpc}, which allows it to measure the execution of the tracee enclaves. It further makes them inaccessible to untrusted software and the enclave itself.
During context switches out of a tracee enclave, it stores the collected \ac{hpc} data in a buffer that the tracer can then access during verification.
Using this approach, malicious code in the enclave cannot interfere with the measurement data. We explain our approach in detail in \Cref{sec:hpc_for_cfa}.
Alternatively, it should be straightforward to integrate hardware-based tracking approaches like~\cite{lofat2017}, where a modified CPU logs information about certain control flow events.
The hardware component could either write directly to the \ac{sm} manged buffer or trigger an interrupt to instruct the \ac{sm} to copy the data from an internal buffer.

Our approach could also work with software instrumentation, although with reduced security guarantees.
This is due to the fact that the measurement data is captured inside the enclave, thus allowing an attacker to potentially modify the measurements before they are sent to the \ac{sm}, if they gain full control over the tracee.

\subsection{Verification Workflow}
\begin{figure}
    \centering
    \includegraphics[width=0.45\textwidth]{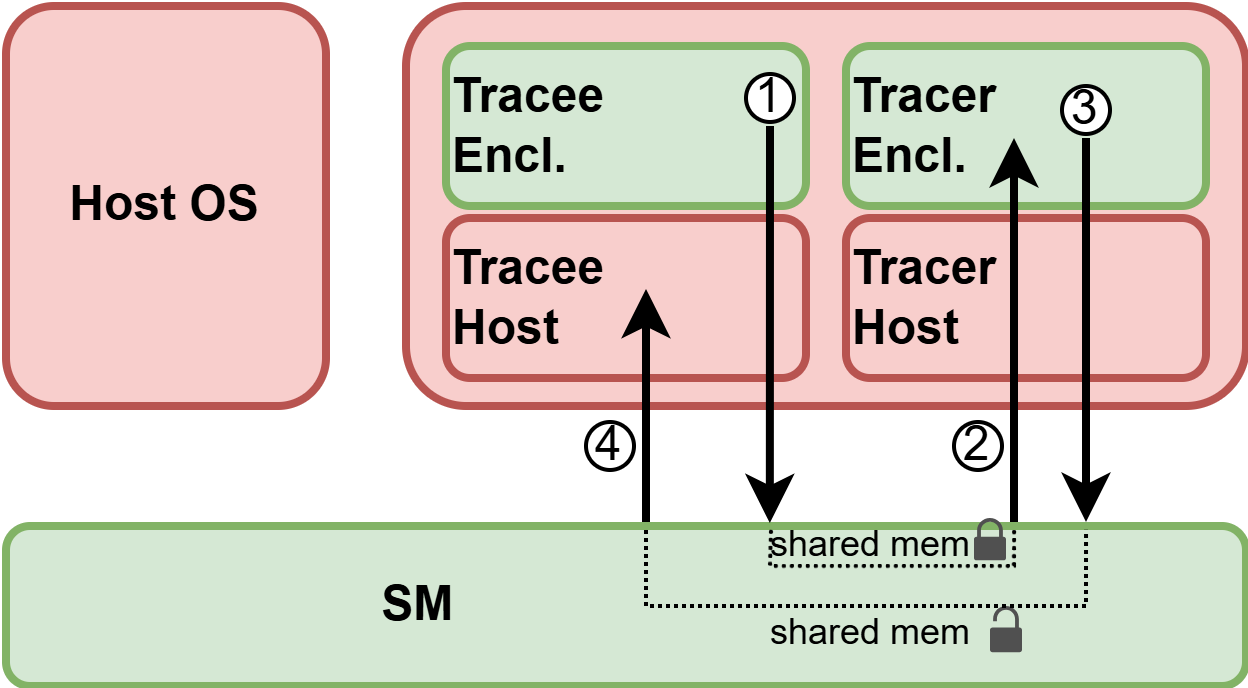}
    \caption{Verification workflow for a tracee tracer tandem. When the tracee performs an \ecall, the \ac{sm} ensures that the shared memory region is only accessible after the control flow has been verified by the tracer.}
    \label{fig:concept-verification-flow}
\end{figure}

\Cref{fig:concept-verification-flow} shows the workflow for the verification.
During startup, the tracee and tracer enclave perform mutual attestation to establish trust.
As explained above, this is necessary because only a tracer enclave trusted by the tracee instance can be allowed to access its control flow data.
While the tracee enclave is running, it is in the \texttt{GATHERING} state and its control flow is monitored by one of the mechanisms described in the previous section.
Upon encountering and executing an \ecall instruction, the \ac{sm} takes over (1) and sets the tracee's verification state to \texttt{PENDING}, signaling that verification from the tracer is required, before unlocking shared memory regions.
Next (2), the \ac{sm} passes control to the tracer enclave to initiate the verification process.

After being invoked, the tracer requests the control flow data from the \ac{sm}. It then performs the verification and (3) returns control to the \ac{sm}.
If the verification was successful, the \ac{sm} unblocks the shared memory region (4) and invokes the host application of the tracee enclave which can now handle the \ecall requested by the enclave in the first step.
If the verification fails (not depicted) the \ac{sm} stops the tracee's execution, keeping the shared memory area blocked to prevent any sensitive data from being leaked. 

Apart from its connection to a tracer enclave, the tracee enclave provides all functionalities that any other enclave does. 
The same does not necessarily apply to the tracer enclave, since, depending on the implementation, it does not need a host application, as it only has to interact with the \ac{sm}.

\section{Attestation Mechanism}\label{sec:hpc_for_cfa}
After having established a concept to achieve general runtime attestation in \ac{tee} scenarios, we continue with the concrete case of \acp{hpc} measurements as control flow traces.

On a high level, our approach works as follows.
First, we use static code analysis to obtain the \ac{cfg} of a program where each node corresponds to a \ac{bb}.
Additionally, each node stores the influence that the corresponding \ac{bb} has on the \acp{hpc}.
At runtime, we observe the execution of the the enclave via \acp{hpc}.
Every time the enclave exits, thus passing control to the \ac{sm}, the monitor stores a snapshot of the \ac{hpc} values together with the current instruction pointer of the enclave.
To verify the control flow of the program between two exits, the algorithm identifies the nodes in the \ac{cfg} corresponding to the instruction pointer values and decides whether there is a path between them that produces the observed performance counter values. 
Note that our concept of a control flow violation depends solely on the \ac{cfg} of a program. 
If, for example, we have a loop on the \ac{cfg}, we consider any number of loop iterations to be valid, even if the underlying program code uses a static for-loop that is repeated exactly $16$ times. 
If, however, the for-loop is unrolled by the compiler, the fixed number of loop iterations is baked into the \ac{cfg} making deviations a control flow violation.

This approach stands in stark contrast to software- or hardware-based instrumentation that can capture both the source and the target address of all control flow transfers, reducing control flow verification to checking whether the encoded path exists in the \ac{cfg}.
In the worst case, we have to evaluate each path between the two nodes in order to verify the control flow.
However, evaluating the control flow only at static control flow transfers to the security monitor that happen due to \ecalls, has the advantage that these paths can be precomputed, thus reducing the verification time.
We propose two optimizations for this basic approach; one reduces the search space when handling returns from function calls, the other allows efficient handling of loops.

\subsection{Call Stack}
Programs often call the same function from multiple locations.
Thus, the nodes that correspond to \acp{bb} with return instructions have many edges to various far-spread locations in the \ac{cfg}.
This quickly makes searching for all paths between two nodes intractable and also increases the chance of false attestations.
By keeping track of the call stack of the program, we can drastically reduce the number of return edges that we need to consider.

Parsing the enclave's calls stack from its stack memory during exits to the \ac{sm} is very challenging, because stack memory does not have a well-defined layout.
While some architectures like RISC-V, do make use of a link register, the return address is still stored to the stack as soon as function calls nest, which is prevalent in most programs.
Instead, we use an inductive approach. From the initially empty call stack, we add and remove entries to and from the call stack as we verify the individual control flow segments.

To precompute the allowed call stacks for each return edge, we add an additional step to the \ac{cfg} generation.
We build a function level \ac{cfg} of the program and assign possible call stacks to each function based on its reachability.

If recursion is utilized, the reachability of certain functions becomes more complex and their number of possible call stacks can become infinitely large. As handling recursion correctly would add a lot of complexity, we made the decision to leave this open for future work and allow no recursion in our current work. Note that this does not limit the capabilities of the attested enclaves, as every recursion can be rewritten as an iteration.

\subsection{Loops}\label{sec:loops}
\begin{figure}
    \centering
    \includegraphics[width=0.3\textwidth]{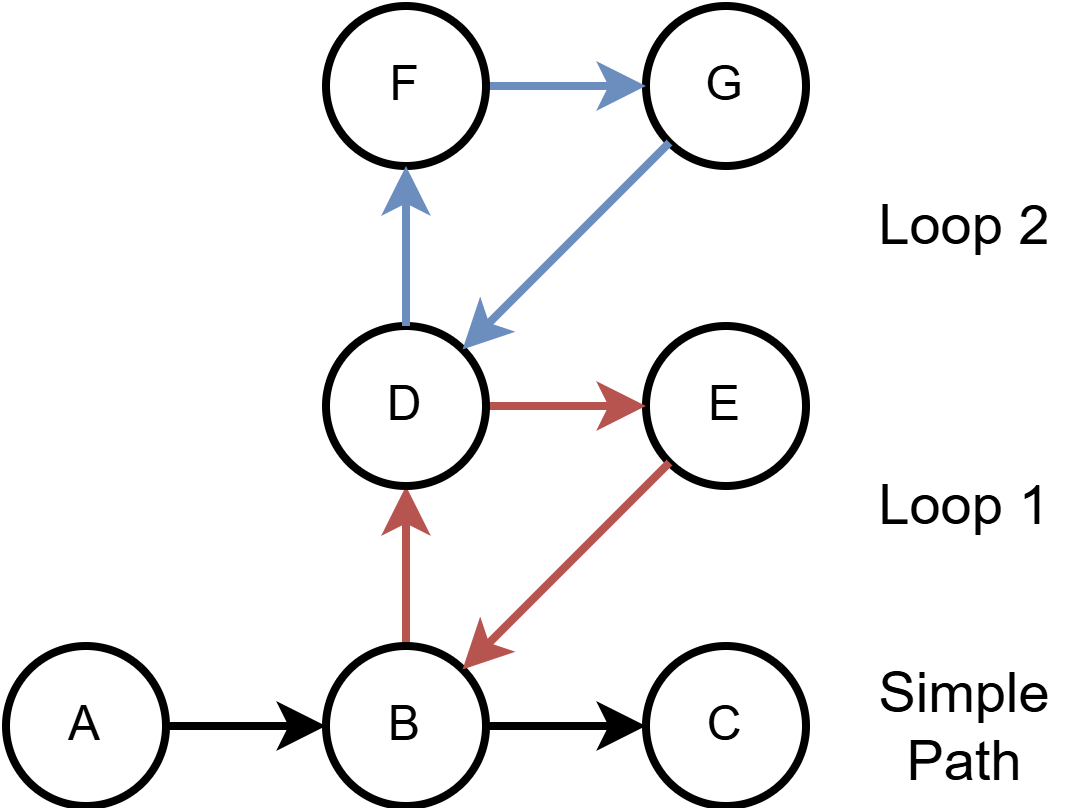}
    \caption{Example section of a \ac{cfg} with a simple path and transitively connected loops.}
    \label{fig:simple-paths}
\end{figure}

Similarly to recursion, loops pose a special challenge for control flow verification, the reason being that the \ac{cfg} allows an arbitrary number of loop iterations. Even if the number of iterations is fixed in the source code, the \ac{cfg} does not retain this information (unless the loop is unrolled during compilation). 
An excerpt of a \ac{cfg} is depicted in \Cref{fig:simple-paths}. If we are, for example, at node B, it is always allowed proceed to either node D, entering loop 1, or to node C, continuing on the simple path. 
In line with related work \cite{cflat2016,lofat2017}, we therefore consider arbitrary numbers of loop iterations to be valid. 

To prevent loops from breaking our scheme, we initially search for all loop-free \emph{simple paths} between two given nodes in the \ac{cfg}.
Afterwards, we enrich each simple path with the transitive closure of all loops that it touches.
In the example shown in \Cref{fig:simple-paths}, the nodes $A\rightarrow B \rightarrow C$ make up the simple path and $B\rightarrow D \rightarrow E \rightarrow B$ and $D \rightarrow F \rightarrow G \rightarrow D$ are the loops connected to the simple path.
Any number of loop iterations that, combined with the simple path, produce the measured performance counter values is accepted as valid. To simplify the verification algorithm, we do not enforce that inner loops, such as Loop 1 in \Cref{fig:simple-paths}, are taken at least once for the outer loop (Loop 2) to be available.
This simplification will never cause any valid control flow to be rejected by the verification; however, it could allow malicious control flows to be falsely verified. However, our evaluation (\Cref{sec:evaluation}) reveals how the reliability of the verification can be increased by other means.

\subsection{Formal Verification Model}\label{sec:formal-model}
Deciding whether a path in combination with its connected loops, is a challenge in itself. The effect of a simple path on the \acp{hpc} is easy to determine, one must simply sum up the effect of all \acp{bb} od the path. However, each linear combination of loops needs to be considered to determine whether a measured set of \ac{hpc} values is valid. Simply testing all possibilities is of course infeasible, let us therefore look at the problem more formally.

Each simple path and each loop can be modeled as a vector $v_i$ that encodes their effect on the performance counters.
Given an \ac{hpc} measurement $m$, we need to find scalars $x_i$ such that $m = p + x_0 \cdot v_0 + \dots x_n \cdot v_n$. The vector $p$ encodes the simple path and is thus not multiplied by a scalar.
Further, the scalars $x_i$ encode loop iteration counts, therefore these are restricted to positive integers.
The structure encoded by this equation is referred to as an integer cone. A close relation to lattices is given by definition, as allowing negative scalars would result in the definition of lattices.

\paragraph{ILP Formulation} 
To algorithmically solve the integer cone membership problem, we model it as an \ac{ilp} problem:
\begin{lstlisting}[mathescape]
maximise:   $x_0\cdot v_0 + x_1\cdot v_1 + ... + x_n\cdot v_n$
subject to: $x_0\cdot v_0 + x_1\cdot v_1 + ... + x_n\cdot v_n \leq$ m-p
            $x_0,...,x_n                   \geq 0$
\end{lstlisting}
The first constraint makes sure that the solution does not exceed the target vector in any dimension. The second one constrains the scalars to allow only positive values. The maximize function is designed to find a solution that is as close to the measurement as possible.
If a valid combination of loops can produce the measurement, it follows that a solution that reaches exactly the target \ac{hpc} values (i.e. $m-p$) exists.
If we therefore find such a solution, we also know that a valid combination of loops can produce the measured values. If, however, the result is less than the target result in even one dimension, we conclude that there is no way that a combination of the given simple loops can result in the target \ac{hpc} values.

\subsection{HPC Selection}\label{sec:hpc-selection}
Naturally, this verification algorithm only works if the selected performance counters behave deterministically.
This disqualifies counters that, e.g., track cache misses or microarchitectural stalls.
However, this still leaves a large number of viable performance counters.
In \Cref{sec:implementation-in-keystone}, we provide additional details on how we profiled the performance monitoring counters on RISC-V to build a mapping from instructions to their influence on the \acp{hpc}.

Modern processors provide numerous events available for counting, but the number of counter registers available is usually quite limited. While the algorithms presented in this work are able to handle an arbitrary number of \acp{hpc}, it is intuitive that the amount of counters available has an impact on the reliability of the verification.
Additionally, the value distribution of the measured events over the executed program plays a crucial role.
As described in \Cref{sec:formal-model}, the accepted performance counter values for a simple path with $n$ loops are equal to the integer cone generated by the loop vectors $v_1$ to $v_n$.
The reliability of our verification approach decreases as the density of the integer cone increases.
In the extreme case, where the cone is equal to $\left( \mathbb{Z}_{>0} \right)^n$, any measurement would be accepted as valid.
Thus, we aim to select the performance counters such that given the value distribution of the program we minimize the density of the integer cones generated by the simple path segments.
As the density of the integer cone itself is not trivial to compute, we use the density of the corresponding lattice as a proxy.
This is based on the fact that the density of the cone converges towards the density of the corresponding lattice the larger the scalars become \cite{intcone1,intcone2}.
We perform an empirical study of the influence of both the number and the value distribution of the performance counters used in \Cref{sec:eval:instrumentation}.

\section{Implementation in Keystone}\label{sec:implementation-in-keystone}
\begin{figure}
    \centering
    \includegraphics[width=0.45\textwidth]{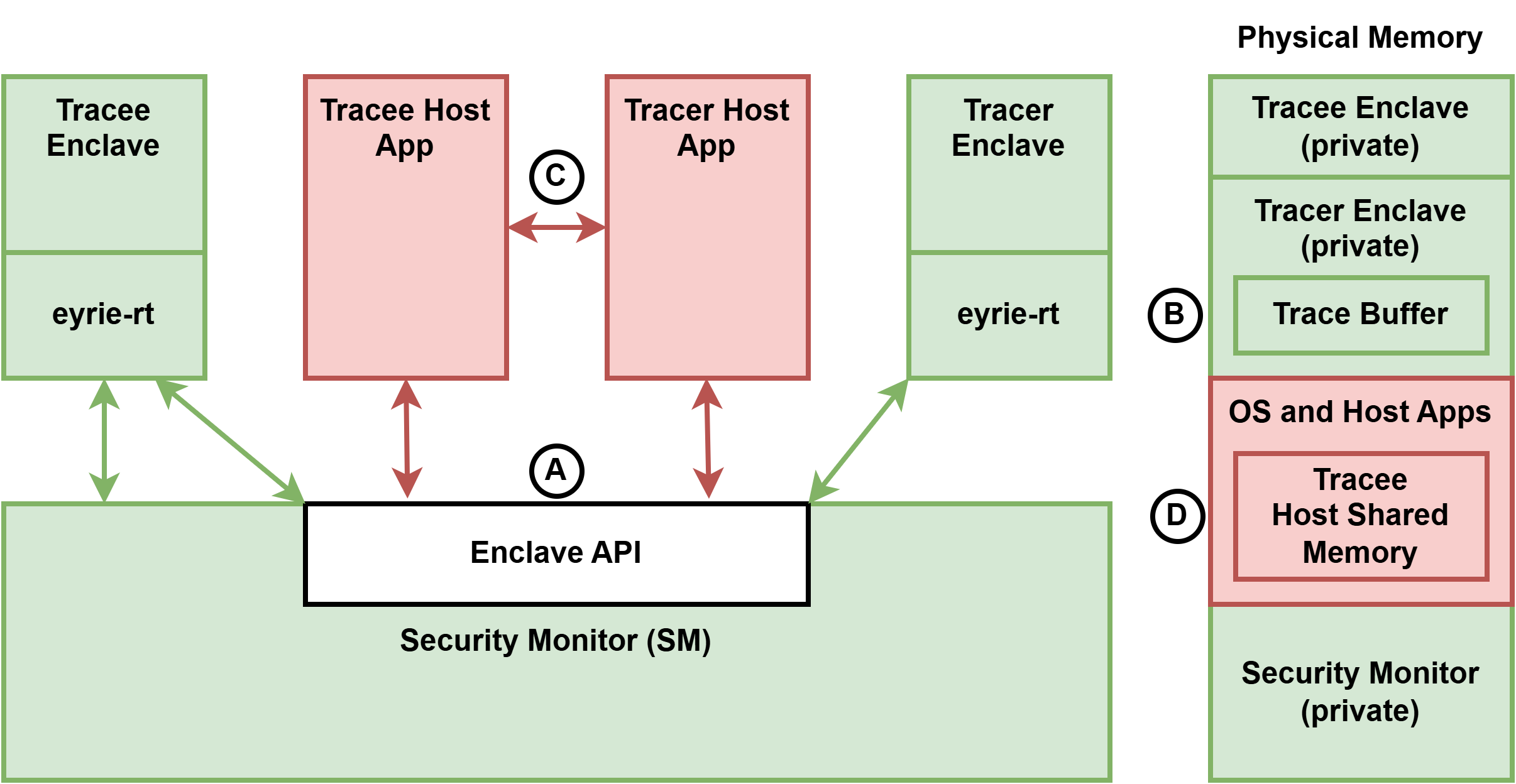}
    \caption{Overview of our tracer tracee architecture implementation in Keystone.
    We modified the enclave API (A) to add support for different roles and the transfer of control flow information to a tracer controlled buffer (B). The tracer and tracee enclaves communicate via \ac{ipc} (C) and the shared memory of the tracee enclave with the host (D) is only unlocked after the control flow verification succeeded.}
    \label{fig:tracer-tracee-components}
\end{figure}

In this section, we describe our proof of concept implementation of the tracee/tracer concept from \Cref{sec:system-design}, as well as the \ac{hpc}-based control flow tracing from \Cref{sec:hpc_for_cfa} for the RISC-V Keystone \ac{tee}.
\Cref{fig:tracer-tracee-components} shows an overview of all components involved.

\subsection{SM Modifications}
To implement the concept of tracer and tracee enclaves, some additional metadata needs to be stored for each enclave and Keystone's API (A in \Cref{fig:tracer-tracee-components}) needs to be extended.
We modify Keystone's central enclave data structure to include some static configuration information that is part of the attestation report, as well as some dynamic metadata that may change during the lifetime of an enclave. The former can, for example, be used to verify that a valid tracer has been attached to a tracee enclave, while the the latter is required to facilitate the verification workflow.

During enclave creation, we add a role to the static information for each enclave, this be \textit{TRACEE}, \textit{TRACER}, or \textit{NONE}, where the latter marks a regular enclave, without \ac{cfa} support. Both tracee and tracer enclaves also store a reference to their counterparts.
Additionally, each enclave was extended with a field to specify the expected attestation value of the associated tracer enclave. Since this field is only relevant for enclaves with the \textit{TRACEE} role, it is required to be zeroed for all other roles.
For tracee enclaves, the dynamic metadata consists of the current verification state, as well as the performance counter readings and the instruction pointer during the least recent exit. 

To facilitate the verification workflow we added two API calls that are only available to tracer enclaves.
The \textit{attach\_as\_tracer} API call, connects a tracer to its tracee enclave. It first checks whether the calling enclave is allowed to trace the specified tracee enclave by comparing the static attestation value of the calling enclave with the reference value specified by the tracee. In addition, the tracer needs to provide a buffer residing inside its private memory range, which is used by the \ac{sm} to share the control flow measurements of the tracee with the tracer  (B in \Cref{fig:tracer-tracee-components}).
If these checks pass, the \ac{sm} stores a reference to their respective counterpart in both the tracer and tracee enclaves and marks the tracee as runnable.
The \textit{set\_cfa\_verification\_state} API call, allows a tracer to update the verification state of its tracee. 
The host OS is blocked from accessing the shared memory region of the tracee enclave as soon as it starts executing and access is only granted after the tracer has successfully verified the tracee's control flow.
Accessibility to the shared memory region is controlled via a dedicated \ac{pmp} entry.

\subsection{Verification Workflow}
Following the enclave model, both the tracer and the tracee are split into an untrusted host application and the trusted enclave which are referred to as host app and \eapp in Keystone.
Both tracer and tracee, therefore run independently and in separate processes. Communication between the two parties is implemented via an \ac{ipc} mechanism  (C in \Cref{fig:tracer-tracee-components}) between the two host apps.
In detail, the communication between the two parties is structured as follows.

\textbf{Initialization:}
The tracee host application is started first.
During initialization, the expected attestation value of its tracer is passed to the \ac{sm} and a POSIX handle for a shared memory object is created. To avoid confusion with the shared memory region, used by the \eapp to communicate with its host app, we hence refer to this memory as \ac{ipc} memory in the following. Afterwards, the tracee waits for its tracer.
The tracer is then started with a handle to the \ac{ipc} memory and the id of its tracee.
During its initialization, it calls the \textit{attach\_as\_tracer} API to register with the \ac{sm} and then notifies the tracee's host app via the \ac{ipc} memory. Afterwards, it sleeps, waiting for verification requests.

\textbf{Runtime:}
After being notified that the initialization is complete, the tracee host app calls the \ac{sm} to start executing its \eapp.
When the \eapp performs an \ecall, the \ac{sm} returns control to the tracee host application as would be the case for any default enclave, however, it keeps the shared memory region blocked.
Next, the tracee host application wakes up the tracer host application via their \ac{ipc} memory.
That in turn triggers the tracer \eapp to perform the verification and communicate the outcome to the \ac{sm} via the \textit{set\_cfa\_verification\_state} API call.
If the verification was successful, the \ac{sm} then unlocks the shared memory region.
Afterwards, the tracer \eapp switches back to its host application, which again uses the \textit{ipc} memory to inform the tracee host application that the verification has been completed.
Finally, the tracee host application can access the shared memory region of its \eapp, to handle the original \ecall~(D in \Cref{fig:tracer-tracee-components}.

We made the deliberate decision not to task the \ac{sm} with scheduling the execution of the tracer, to keep the \ac{tcb} small and to not interfere with the regular OS scheduling logic.
The downside of this design is additional overhead due to switching between two processes.
We discuss alternative designs in \Cref{sec:discussion}.

\subsection{Generating Control Flow Traces}
In this section we explain our configuration of the \acp{hpc} to trace the execution of enclaves.

As documented in~\cite[Sec 3.1.10]{riscvPrivileged} the performance counter subsystem on RISC-V is configured using a set of M-mode \acp{csr}.
Each counter $x$ can be configured via the \textit{mhpmcounterX} register, while the \textit{mhpmeventX} contains the current value of the configured counter event.
The RISC-V specification mandates $29$ counter registers but implementers may choose to hardwire them to zero.
The semantics of the configuration registers and the supported events are also implementation-specific.
In addition, there are two fixed, mandatory counters, \textit{mcycle} and \textit{minstret}, that count cycles and retired instructions.
The \textit{mcountinhibit} register can be used to start and stop individual counters without changing their configuration.
The \textit{mhpmcounterX} and \textit{mcountinhibit} are only accessible in M-mode and thus exclusive to the trusted \ac{sm}.
Lower-privileged software is required to use a software defined API of the code running in M-mode, to configure the performance counters.
The accessibility of \textit{mhpmeventX}, \textit{mcycle} and \textit{minstret} in lower privilege modes can be controlled via the \textit{mcounteren} register.
The combination of these mechanisms allows us to ensure that only the \ac{sm} can configure the performance counters, therefore protecting them from untrusted components.

Additionally, we need to restrict the performance counters to only count the events of a specific enclave.
Since the RISC-V specification does not provide any mechanism to count only the events of a certain process, we need to manually implement this behavior.
We use the \textit{mcountinhibit} register to configure to enable the counters shortly before the context switch to the enclave and to disable them as soon as execution returns to the \ac{sm}, directly inside the M-mode assembly trap handler. 
We then store the observed difference in \acp{hpc} in the context of the respective enclave.
Using the \textit{mcountinhibit} register conveniently allows us to enable and disable all counters at once using a single \ac{csr} write.  
As we still need to free up a scratch register before we can modify the \textit{mcountinhibit} register as well as perform context switches from and to the \eapp, the counters are influenced by a small number of instructions that do not belong to the traced code. However, they are fixed which means that we can easily revert their footprint on the counters.

The \ac{hpc} measurement logic is further complicated by the fact that the trap handler can also be triggered while we operate in M-mode.
While disabling all performance counters at the start of the handler is idempotent and thus unproblematic, we may only enable performance counters when we triggered the handler from an enclave. To this end, we check the \textit{MPP} field in the \textit{mstatus} \ac{csr} which reveals the privilege mode at which the trap event occurred. We carefully crafted the assembly code to prevent the branching logic from influencing the performance counter values in a dynamic manner.

In principle we could still enable regular software to use performance monitoring counters, by extending the context switching logic to also store/restore their values and configuration before entering an enclave, however, we decided against this in our proof-of-concept implementation as this would add unnecessary complexity.

\subsection{HPC Prediction}
For our verification approach, we need to be able to compute the impact, that a given code sequence has on the the \acp{hpc} without the need to execute said code.
In a first line of experiments we analyze which \acp{hpc} behave deterministically, meaning they always report the same values for the same instruction sequence. Next, we conducted experiments to evaluate the behavior of different \ac{hpc} events with respect to executed instructions.

We then created a list of all RISC-V instructions supported by the VisionFive2 SoC. Subsequently, we created a sample enclave containing each of these instructions in short sections fenced by \ecall instructions. This allowed us to create an exact mapping that shows which \ac{hpc} event is triggered by which instructions. This mapping is the heart of our \ac{hpc} prediction. Based on this mapping we can simply parse all instructions in a given \ac{bb}, which allows us to add up the effect that this \ac{bb} will have on the \ac{hpc}.

\subsection{\acs{cfg} Generation}
\label{sec:cfg}
Attesting the control flow of an application requires knowledge about all allowed paths through the program. The usual way to gain this information is to create the \ac{cfg}, in which all valid transitions between \acp{bb} are encoded as edges. The automatic generation of \acp{cfg} is a complex task that, to the best of our knowledge, has no perfect solution.
The main hurdle for \ac{cfg} generation is \emph{indirect branching} where the target of a control flow transfer instruction is not hard coded but dynamically computed at runtime~\cite{DBLP:conf/uss/TsangAJSMH24}.
Examples where this behavior is used are the implementation of features like function pointers in C or virtual functions in C++.

In this work we use Ghidra to perform a static code analysis on the binary of our tracee enclave.
Its API allows us to create a plugin that iterates over all \acp{bb} and their control flow transitions to generate the \ac{cfg} for the whole enclave. Next, we add the \ac{hpc} predictions to each \ac{bb} and export the result into a file that is then used for the attestation workflow.

We found that Ghidra does not reliably detect all targets of indirect branches, i.e. if function pointers are used. Our basic implementation therefore does not support the verification of e.g. dynamic dispatch.
However, to show that \hpcfa also works for these scenarios, we have conducted additional experiments. For these we compile a tracee enclave with a custom LLVM compiler plugin, which enables us to predict all possible targets for indirect branches.
Enriching our Ghidra-generated \ac{cfg} with these branch targets ensures that the \ac{cfg} is complete, allowing us to perform our usual attestation.

\subsection{Attestation}
We have split the implementation for the verification of the traces into two steps.
The first step is a Rust program that performs a one-time preprocessing of the \ac{cfg} to compute all possible paths between \ecalls (including simple paths and simple loops), while adhering to valid call stacks. Note, that this does not require the execution and measurement of enclave code.
The second step is a Python program that uses this information to perform the actual verification algorithm described in ~\Cref{sec:hpc_for_cfa}.
We use the \ac{cbcsolver}\footnote{\url{https://github.com/coin-or/Cbc} (Accessed 2025.04.15)} to solve our \ac{ilp} problem. As it was not possible to specify the problem with vectors as described in \Cref{sec:formal-model}, the first constraint had to be expanded into one constraint per vector dimension. Similarly, the objective function needed to be expanded, summing up the resulting objective functions.

For our proof-of-concept implementation, we have not yet integrated the verification into the tracer enclave.
Instead, the tracer enclave writes the trace data into a file which is then analyze on a separate system.
This is mainly motivated by the fact that Keystone enclaves do not easily allow the usage of high-level programming languages such as Python restricting the ability to quickly prototype different approaches.
For an end-to-end deployment, our current Python proof-of-concept would need to be ported to C.

\section{Evaluation}\label{sec:evaluation}
In this section we evaluate both the reliability and performance of our approach.
We run Keystone on a StarFive VisionFive2 SoC which contains four SiFive U74 CPU cores.
The proof-of-concept implementation of the verification runs on an Intel Xeon Gold 6342 CPU.
We first introduce the analyzed applications as well as our performance and reliability metrics.
In \Cref{sec:eval:vf2} to \Cref{sec:eval:instrumentation} we apply our metrics to the base version of our approach before exploring two variations to improve its reliability.

\subsection{Analyzed Applications}~\label{sec:anaylzed-application}
For our experiments, the tracee needs to perform realistic computations for which we can perform an attestation.
However, the restricted programming environment offered by Keystone enclaves, which lack full POSIX compatibility, drastically limits the number of programs that can be run without major porting efforts. For all of our example enclaves we use Keystone's default runtime --- the eyrie runtime \cite{DBLP:conf/eurosys/LeeKSAS20}. 
Further, we avoid dealing with two address spaces and, therefore, start our attestation after the runtime has set up its virtual address space. Further, we restrict our payloads to always run in S-Mode, thus avoiding context switches between S-Mode and U-Mode. 

The first workload we analyze is a simple "Hello, World!" program (\hellotracee), where the \eapp simply sends a string to be printed to its host app. This is a useful program to start with as it mainly consist of runtime code, which is also used by all other enclaves. 

This also revealed that the runtime makes use of recursive functions. As discussed above, our call stack logic is unable to handle recursion. We therefore decided to include one measurement point right before and one right after such recursive functions. When encountering a segment that starts after the first and ends with the second inserted \ecall during verification or evaluation, we can simply skip that segment and continue with the next one.

Next, we have our main workload (\tweetnacltracee), based on the minimal tweetnacl\footnote{\url{https://tweetnacl.cr.yp.to/software.html}} cryptographic library. 
Despite its small size, it still performs computations relevant for real-world workloads and is well suited to be used within the restricted environment of enclaves.
Upon execution, our evaluation tracee enclave creates an asymmetric key pair, signs the message \emph{"Hello, world!"}, sends the signed message to the host app, then verifies the signature on the message, and finally sends the verification result to the host application before terminating.
Lastly, we have an enclave that contains a minimal code example making use of dynamic dispatch (\dyndistracee).

\subsection{Reliability Metrics}
\label{sec:reliability-metrics}
We start by explaining the metrics that we use to determine the reliability of our approach. 
With these metrics, we want to analyze \hpcfa's efficacy in detecting code reuse attacks, such as \ac{rop}~\cite{DBLP:journals/tissec/RoemerBSS12} or \ac{jop}~\cite{DBLP:conf/ccs/BletschJFL11} attacks. 

\hpcfa is designed to correctly attest all valid control flows and never cause any false rejections of valid control flows. Therefore we only test cases in which the control flow has been violated, as we want to determine how many of these erroneously pass the attestation.
Further, we aim for scenarios in which an attacker only modifies only small parts of an otherwise valid control flow because we consider these to simulate the worst case. We expect that in more realistic cases an attacker would require significantly larger changes, which in turn should be easier to detect.

To simulate such small modifications to a valid control flow, we trace a whole run of an enclave in QEMU\footnote{\url{https://www.qemu.org/} (Accessed 2025-11-20)}, recording all basic blocks that were executed. This trace can be split into multiple sections separated by \ecalls that are verified and evaluated individually. Note, that we always evaluate all sections for a given enclave when reporting reliability results.

We evaluate the following modifications on the basic block level:
\begin{itemize}
    \item \emph{Replace block}: Replace a \ac{bb} from the original trace with a different one.
    
    \item \emph{Replace with unique block}: Like the previous one but we ensure that the added block has a unique \ac{hpc} delta. This primarily excludes very small \acp{bb}.

    \item \emph{Insert unique block}. Add a block with a unique \acp{hpc} delta to the trace.

    \item \emph{Remove block}: Remove a \ac{bb} from the original trace, simulating an attacker that skips, .e.g, an if branch in program.
\end{itemize}
Furthermore, we also analyze more fine-grained changes to the control flow that mimic a scenario, where an attacker has, for example, access to an exploit that enables code injection.
\begin{itemize}
    \item  \textit{Random change}:  We introduce random changes to the measured performance counter values in the range of $\pm10\%$.
\end{itemize}

Our basic reliability score is the relative amount of modified segments that fail the attestation, i.e. for which the verification is working.
Since the verification occurs at each \ecall, we analyze all of the described modifications for each \ecall to \ecall segment of a full trace.
Because \ecalls are usually not distributed evenly within one trace, the length of the individual segments varies drastically, ranging from less than $100$ instructions up to over $100$ million instructions.
Thus, we use two metrics. 
The first uniformly averages the reliability scores over all segments. The second weighs each segment by its original instruction count. This second metric compensates for cases in which there could be many short segments with perfect reliability and one big segment with $0$ reliability which would in total report a very high reliability value according to the first metric, even though most of the program's execution was vulnerable.

For each \enquote{replace} experiment we aim to compute the average over \num{1000} repetitions per segment while the other experiments aimed to be averaged over \num{100} repetitions per segment. Note that we prevent any two repetitions from being the same, meaning that there are cases in which we compute the average over fewer repetitions. For example, if we have a segment consisting of $8$ \acp{bb}, we can only remove $8$ different \acp{bb}, therefore we base the average on these $8$ experiments.
When removing blocks from the trace, we made sure that the resulting path is not itself valid. This would, for example, be the case when removing a block that corresponds to the body of an if-statement without an else case. Similarly, we ensured that inserting a block into the trace did not directly create another valid control flow.

\subsection{Performance Metrics}
The other aspect we evaluate is the performance overhead. We look at two different aspects: first, the performance overhead introduced by the context switches due to the tracer/tracee design, and second, the raw overhead introduced by the verification algorithm.

The overhead introduced by the tracer/tracee design is measured on the VisionFive2 board using the \texttt{time} console command. As a baseline, we measure the time that the tracee enclave needs without any tracer, which is obtained by starting it as a normal enclave.

For the verification, we time the execution of each segment inside our evaluation script. This excludes some initial setup that takes place before the segment verification and only needs to be done once.

\subsection{Results: Basic Approach}
\label{sec:eval:vf2}
For the evaluation, we focus on the \hellotracee and \tweetnacltracee applications described in ~\Cref{sec:anaylzed-application}. 
The \dyndistracee only contains a minimal example that makes use of dynamic dispatch via function pointers. Performing some steps of the evaluation on it revealed that the results do not create added value, which is the reason we omit them here. However, it is important to note that the presence of dynamic dispatch does not in any way influence the verification algorithms presented here, as long as the correct edges are present in the \ac{cfg}.

In this first evaluation, we run our tracee applications on the StarFive VisionFive2 SoC.
In addition to hardcoded instruction and cycle performance counters, the board only provides two configurable performance counter registers. We are using the following counter events:
\begin{itemize}
    \item Fixed Counter: Instructions Retired
    \item Counter 0: conditional branch instructions retired + \texttt{jal}-instructions retired + \texttt{jalr}-instructions retired
    \item Counter 1: integer loads retired
\end{itemize}
Note that counter 0 is configured to count three events at the same time.

The \hellotracee performs only three \ecalls, of which the first happens right after the virtual address space has been set up. This also marks the start of our verification algorithm. 
The \tweetnacltracee performs significantly more computations, but only performs seven \ecalls.
With both enclaves executing hundreds of thousands of instructions, there are long sequences of code executed between these measurement points.
Our analysis revealed that this leads to an intractable number of paths between the \ecalls, which made our preprocessing approach infeasible, because the resulting file, containing the pre-computed paths, had a size of multiple Gigabytes. 
We therefore added $16$ additional \ecalls between key computation steps of the eyrie runtime, which apply to both of our enclaves. For the \tweetnacltracee we added another $8$ \ecalls to its payload code. 
Note that none of these \ecalls have any effect other than passing the control back to the host OS, thus triggering verifications. The addition of these measurement points drastically reduces the number of pre-computed paths making the verification tractable.

\paragraph{Reliability}
The results of our reliability evaluation are found in \Cref{tab:reliability-results} and show that this approach in its basic form reports different reliabilities for our two enclaves. 
For the \hellotracee enclave most deviations from the control flow are detected, while the results for the \tweetnacltracee enclave show that \hpcfa provides barely any protection.
The trace of the whole enclaves is divided into a total of $21$ respectively $30$ segments of different sizes.
For shorter segments, we observe high reliability in detecting deviations, while longer segments tend to have very low reliability.
This also explains the different reliability results for the two enclaves, because the trace of the \hellotracee enclave, which is mostly the runtime, is divided into shorter segments, while the \tweetnacltracee also contains these short segments, but additionally a few very long segments.
This further explains the strong difference between our two metrics for the \tweetnacltracee enclave, because it performs the vast majority of its computations inside these few long segments.
Based on this observation, we evaluate the impact of restricting the length of each measurement segment in \Cref{sec:eval:instrumentation}.

\begin{table*}[t]
    \centering
    \begin{tabular}{c|c|c|c|c|c|c}
Experiment & Enclave & Random Change & Remove Block  & Replace Block & Replace with Unique B. & Insert Unique B.\\
\hline
Basic Scenario & \hellotracee & 0.955, 0.986 & 0.923, 0.973 & 0.924, 0.971 & 0.932, 0.975 & 0.930, 0.958\\
Basic Scenario & \tweetnacltracee & 0.803, 0.000 & 0.741, 0.000 & 0.743, 0.000 & 0.753, 0.000 & 0.744, 0.000\\
Added Counters & \hellotracee & \textbf{1.000, 1.000}   & \textbf{0.991, 0.997} & \textbf{0.996, 0,999}     & 0,994, 0.998     & \textbf{0.994, 0.998}  \\
Added Counters & \tweetnacltracee & \textbf{1.000, 1.000}   & 0.886, 0,000     & 0,900, 0.092     & 0.906, 0.128   & 0.900, 0.110\\
Added \ecalls & \hellotracee & \textbf{1.000, 1.000}   & \textbf{0.991, 0.997}  & \textbf{0.996, 0.999}  & \textbf{0.995, 0.998}   & \textbf{0.994, 0.998}\\
Added \ecalls & \tweetnacltracee & \textbf{1.000, 1.000}   & \textbf{1.000 1.000}  & \textbf{1.000, 1.000}  & \textbf{1.000, 1.000}   & \textbf{1.000, 1.000}\\
    \end{tabular}
    \caption{Reliability evaluation results for all experiments. The first number is the rate at which segments are rejected correctly while the second one takes into account the total number of instructions that were verified to be correct.}
    \label{tab:reliability-results}
\end{table*}

\paragraph{Performance Overhead}
The results of our performance evaluation can be seen in the third column of \Cref{tab:performance results}. 
We observe a significant increase in execution time, when using \hpcfa. However, such an overhead is to be expected, since the execution has to alternate between the tracer and the tracee, which also introduces a lot of additional context switches.
The verification adds another significant amount of overhead that surpasses execution time of the enclave with measurements enabled. 
We discuss the performance impact of \hpcfa in \Cref{sec:discussion} and further discuss how moving certain parts of the tracer enclave into the \ac{sm} could improve performance at the cost of an increased \ac{tcb}.

\begin{table*}[t]
    \centering
    \begin{tabular}{c|c|c|c|c|c}
                            & Enclave & Baseline  & Basic Scenario            & Added counters    & Added \ecalls \\
                            \hline
        Enclave execution   & \hellotracee & $0.41$s   & $1.53$s ($\times3.73$)    & -              & $1.54$s ($\times3.76$)   \\
                            & \tweetnacltracee & $0.81$s   & $1.88$s ($\times2.32$)    & -           & $11.7$s ($\times14.44$)   \\
        Verification        & \hellotracee  & -         & $33$s                 & $11$s            & $11$s \\
                            & \tweetnacltracee  & -         & $1$m $29$s        & $1$m $40$s       & $23$m $13$s
    \end{tabular}
    \caption{Time measurements of the enclave's executions and the verification of the full traces for all experiments. Averaged over 50 measurements.}
    \label{tab:performance results}
\end{table*}

\subsection{Results: Additional Counters} 
\label{sec:eval:additional_counters}
The Visionfive 2 board only allows two configurable performance counters, limiting our approach to measure three counters in total.
In this section, we evaluate how the reliability scales with adding more performance counters.
With our mapping from instructions to their impact on the \acp{hpc}, we can easily simulate any number of concurrent counters for a given trace.
The Visionfive 2 boards offer a total of $17$ deterministic events that we were able to map. 
In this line of experiments we repeat the experiments from the previous section and evaluate how the ability to measure $17$ counters at the same time would influences the reliability and performance.

\paragraph{Reliability}
Our results (\Cref{tab:reliability-results}) show that the additional \acp{hpc} significantly improve the reliability. With over $99\%$ reliability for all deviations for the \hellotracee enclave this experiment shows that \hpcfa works.
The results for the \tweetnacltracee enclave have also been improved, but a reliability of less than $0.1$ in most cases is not enough to justify the performance overhead.

Further investigation reveals that most of the newly added counters are tracking events related to specific instructions, which are effectively absent in our enclaves. 
Therefore, it is not surprising that the addition of these counters did not improve the reliability in a manner that the \tweetnacltracee enclave could be protected.

The fact that random changes are now detected in effectively all cases is likely connected to the \acp{hpc} that are rarely triggered, because changing these can quickly lead to counter states that are not reachable according to the given \ac{cfg}.

\paragraph{Performance Overhead}
The results are shown in the fourth column of \Cref{tab:performance results}. As the additional counters are only simulated, we cannot report the impact on the enclaves execution. We assume that the added overhead would be minimal, because the only required change would be to record more \ac{hpc} values.

The verification with more counters is slightly faster compared to the previous experiment. We suspect that some paths may be easier to discard, if more counters are available, making the \ac{ilp} faster to solve. However, we suspect that this can vary from workload to workload.

\subsection{Results: Additional Measurement Points} \label{sec:eval:instrumentation}
The high reliability in short segments in \Cref{sec:eval:vf2} suggests that additional measurement points can be introduced in order to increase the overall reliability. In this section we evaluate this approach in more detail.

Analyzing the segments with low reliability in more detail shows that they are not only long but also contain a high number of loops.
Applying the formal integer cone model for our verification process, we observe that the vectors corresponding to the set of loops on these paths almost form a basis for the unit integer cone.

Based on these observations, we analyzed the binary and manually inserted additional \ecalls to reduce the number of loops between any two \ecalls.
For the \hellotracee enclave and thus the runtime code, a single additional \ecall was enough to get near perfect reliability. For the \tweetnacltracee, we added another $49$ \ecalls. As some of these \ecalls were inserted into nested loop structures, the number of segments on the trace increases from $30$ to $102619$. 

A side effect of having \ecalls inside loops is that some segments can repeat. Filtering out duplicates, i.e., segments with the same start and end nodes, measured counter values, and initial call stacks, results in a vastly reduced number of segments for the \tweetnacltracee. The total of $102619$ segments can be reduced to only $788$ unique segments, reducing the number of measurements to be verified drastically. 
For the reliability analysis we multiply the results of these segments with their actual frequency, in order to obtain results that are comparable to the previous scenarios.
Note, that we are using our simulated evaluation approach and are still simulating the values of all $17$ counter events. 

\paragraph{Reliability}
The reliability results are depicted in \Cref{tab:reliability-results}.
We see that reducing the size of segments is an effective way to improve the attestation to have near-perfect reliability for all tested control flow deviations.

\paragraph{Performance Overhead}
The performance impact is shown in the last column of \Cref{tab:performance results}. 
This shows that regular measurements of the \acp{hpc} are getting expensive. For the \hellotracee enclave, where we only added one measurement, the overhead is accordingly very small.
However, the \tweetnacltracee, where $102619$ measurements are performed instead of $30$, the execution time of the tracee increases by a factor of $14.44$.
Which results in an overhead that may\textbf{} not be acceptable in many scenarios. However, other use cases that require strong security guarantees may profit from the added value provided by \hpcfa.
Further, it is important to note that our proof-of-concept implementation is not optimized for performance, especially the communication between the tracer and tracee enclave introduces avoidable overhead. 
We decided to design our implementation as we did, because we wanted to evaluate, whether \ac{hpc} predictions can be used to enable \ac{cfa} and our design enables us to more flexibly test out different approaches.
We assume that it is realistic cut the measurement overhead in half with an optimized implementation.

When looking at the time needed for the verification, we also see that this is higher, however, while the other scenarios only lead to $30$ segments, we are now verifying $102619$ segments. This means that the average verification time per segment actually decreased from  $3$s in the first scenario to $13$ms in the last scenario.
Considering the observation that of the $102619$ segments only $788$ differ in their trace, caching attestation results for previously seen segments would further reduce the verification workload in this case, as we would be able to skip more than $99\%$ of the necessary verifications in this case (reducing the verification time to $1$m $9$s). However, this degree of effectiveness will vary based on the tracee enclave, we therefore opted to report the worst case values.

\section{Security Analysis}\label{sec:security-analysis}

Let us start with the security implications that the introduction of tracing has. 
The ability to read \ac{hpc} traces of an enclave poses a threat, as these can be used to infer the control flow path that was taken. It is therefore vital that these traces are protected from potential attackers. Our solution is to generate the traces inside the \ac{sm}, which is part of the \ac{tcb} and only making them available to the attached tracer enclave, which is also inherently trusted by the tracee enclave. This is ensured by defining tracer's hash in the tracee's metadata. Note, that these measurements and traces 
are never exposed to the untrusted host OS or other enclaves.

Next, we look at the tracer enclave. As established before, it has access to the control flow traces, which are security sensitive. We guarantee strong protection of these traces by running the tracer as an enclave inside a \ac{tee}, while severely limiting its interface. All sensitive data processed by the tracer is provided by the \ac{sm}, i.e. a trusted source. The API that the tracer provides to the untrusted system depends on the implementation. In case of our proof-of-concept implementation, the tracer's verification is triggered by the untrusted host app via its API. It does, however, not accept any input from the untrusted system, preventing it from working with invalid or malicious input.
An alternative design could have the \ac{sm} trigger the tracer, which would eliminate the need to provide any API to the untrusted system at all. Further, the tracer has no direct access to its tracee and can therefore not directly interfere with its computations.

Another point that is important to discuss is the influence that a fully compromised enclave has on the generated traces. First note, that an attacker has no influence over the generated measurements, as these are generated by the \ac{sm}. Nonetheless, it is fair to assume, that such an attacker would be able to skip measurement points. 
This could intuitively be considered to make additional measurement points, specifically introduced to increase the reliability, useless. 
This intuitive assumption is incorrect, because the additional measurement points are effective, not simply because the tracee is interrupted more often, but because it keeps the number of valid control flows and therefore expected measurements for each segment small.
As discussed before, it is also not an option for the attacker to never trigger a measurement point again after taking control of the enclave, as this would leave them unable to leak any data from inside the enclave.
This is an important distinction to make when comparing the approach to approaches that rely on instrumentation to generate the measurements, e.g. by tracking taken branches.

\subsection{Covert Channels}
As discussed in our attacker model, we currently consider information leakage through covert channels out of scope for our tracing architecture.
Our enclave design prevents direct interaction with resources outside the enclave before the tracee has been verified.
Nonetheless, an attacker that took control over the tracee enclave could still try to leak information through microarchitectural side-channels.
One way to curb this threat could be to incorporate interrupts, especially timer-based scheduling interrupts, to trigger a control flow verification.
In contrast to the verification during \ecalls this verification could be performed asynchronously in the background.
While this would not prevent the covert channel leakage itself, it would severely limit the time window until detection and thus the amount of data that can be leaked.
One challenge with including interrupts in the verification is that their asynchronous nature precludes the use of precomputed path data.

\section{Discussion}
\label{sec:discussion}
\bheading{Reliability on long segments:}
In \Cref{sec:eval:vf2} we have seen that the \ac{cfa} fails on longer segments, while it works well on shorter segments. The reason for this can be found when looking at the formal verification model in \Cref{sec:formal-model}. On the longer segments, where the evaluation is failing, we have relatively short simple paths, while the biggest share of the measured \ac{hpc} values is the result of various loops. We therefore conclude that the integer cone around our target vector is very dense, causing most simulated attacks to also produce vectors that are part of the integer cone, which then results in false attestations.
Even increasing the dimensionality of the integer cone, which we do in \Cref{sec:eval:additional_counters} does not solve the underlying problem.
To fix the underlying problem, we propose to add additional measurement points in \Cref{sec:eval:instrumentation}. This works on two levels; for one, the number of loops on each segment is reduced, i.e. the integer cone is generated by fewer vectors. And secondly, the measured \ac{hpc} values are smaller, i.e. the candidate vector for the integer cone membership problem has a shorter distance to the origin.

\bheading{Applicability to other TEE solutions:}
While we only provide a proof-of-concept implementation for Keystone on RISC-V, we are confident that our approach can be ported to other platforms, like Intel TDX~\cite{tdx-whitepaper} or Arm CCA~\cite{armCCA}.
Both feature a trusted software layer at the highest privilege level that is responsible for context switching between security contexts and both architectures have \acp{hpc}.
However, only Intel can make changes to the TDX module and there is no hardware yet with ARM CCA support.
The prediction of \ac{hpc} values would need to be repeated for the respective architecture, but this should also be possible even if more tedious due to the more complex instruction set. We were already able to verify that there are \acp{hpc} with deterministic behavior on Intel CPUs.

\bheading{Dynamically choosing counter events:}
During the evaluation, we have increased the number of counters that we are tracing, which is not an issue, when simulating the values. On a real system, providing only a limited number of counter registers, this would not be an option. 
As an alternative, we propose randomize the counter events that are traced for each segment. This would not have the same effect as being able to measure all counters in parallel, however, an attacker would still have to craft attacks, working for all events, in order to avoid detection reliably.

\bheading{More efficient ecalls:}
As detailed in the evaluation, the current design comes with additional overhead when handling \ecalls.
While this enables us to keep the \ac{tcb} small, it might be worthwhile to consider trading off performance vs \ac{tcb}.
Upon encountering an \ecall the control passes to the \ac{sm}, which could then directly trigger the tracer enclave to start the verification. After completion, the \ac{sm} would take over and pass control to the host application. This nets four context switches and only two of those are introduced by the tracer/tracee concept.
Looking at the current implementation we see the following eight context switches: tracee \eapp $\rightarrow$ \ac{sm} $\rightarrow$ tracee host app $\rightarrow$ tracer host app $\rightarrow$ \ac{sm} $\rightarrow$ tracer \eapp $\rightarrow$ \ac{sm} $\rightarrow$ tracer host app $\rightarrow$ tracee host app.

Another way to reduce the footprint of \ecalls is connected to the additional \ecalls we introduce to add measurement points. These are triggering complete context switches to the host OS and the tracer enclave. However, these are dummy \ecalls, that could be caught by the \ac{sm} which only needs to write the trace data to a buffer with intermediate measurements. This could make these interruptions significantly cheaper without undermining the security guarantees, as no communication with the host application happens at these points. 

\bheading{Verification by the TEE:}
Going even further, to reduce the performance impact of the \ac{cfa}, one could move the verification fully inside the \ac{sm}. In this case, the verification would require no additional context switches for each \ecall as the \ac{sm} gains control anyway. These improvements would come at the cost of moving all related computations inside the \ac{tcb}, however, this is a consideration especially if only tracer/tracee enclave pairs are running on a system, which inherently imply trusting the attached tracers.

\bheading{Loop handling:}
Taking a closer look at loop unrolling could help to significantly improve the reliability in some cases, without the need to introduce more measurement points. Especially for cryptographic implementations, where the exact number of iterations is known for a large number of loops, this could be beneficial. However, it is also important to consider that branches in the loop body would be serially connected, which could in turn cause an explosion of viable paths discovered during preprocessing.
Alternatively, it could also be interesting to annotate the loops and use this information during attestation, thus removing the need for further modifications to the binaries. These modifications should be considered in future work. 

\bheading{Dynamic measurement points:}
Another, related approach would be to regularly interrupt the enclave (e.g. based on a timer) to have dynamic measurement points. These could be collected by the \ac{tee} while verification is only required once the tracee \eapp exits. This would also allow starting the verification process in parallel if resources are available. We have decided against this approach to fully enable extensive preprocessing, making the live verification more efficient. However, in future work we intend to look at the use of dynamic measurement points, as this could remove the need for inserting measurement points into the enclave of the binary.
Intel's \ac{pebs}\cite{pebs} allows for hardware-assisted creation of records containing the \ac{hpc} values and instruction counter, upon counter overflow. 
An approach based on dynamic measurement points could profit from hardware implementations like \ac{pebs}, as this would allow the creation of more measurement points with only little overhead and no expensive context switches.

\section{Related Work} \label{sec:relaated-work}
\ac{cfa} is a concept first introduced in C-Flat \cite{cflat2016}. The authors show how the attestation of the control flow enables a remote verifier to check whether a program has been corrupted by e.g. return-oriented programming attacks. 
For the attestation presented in \cite{cflat2016}, a trusted component is needed on the prover (i.e. the monitored device). In the proof-of-concept implementation ARM TrustZone-A is used to create a trust anchor on a potentially corrupted device. The task of this isolated component is to hash and cryptographically secure information about the observed control flow.
The attestation process is implemented as a challenge-response protocol in which a remote verifier provides input to the audited software and later receives proof of the observed control flow. 
The generation of the proof makes use of trampolines that are added to the software's control flow instructions (jumps, returns, etc.) during instrumentation. These trampolines pass the source and destination addresses of the triggering control flow instruction to the trusted component, which uses this information to compute the next value in a hash chain. The last value of this chain makes up the central element of the proof provided to the verifier. Loops are treated in a special way in order to avoid the set of valid hashes for the software to explode. Instead of adding each jump between loop iterations to the hash chain, an individual hash value is computed for each occurrence of a loop, this value as well as the number of iterations observed is then separately added to proof.
For the verification step the control flow graph of the program to be attested as well as its binary are required. Based on these and the input parameters, the verifier checks whether the reported hash value is the correct result for the expected control flow path.

As a follow-up to C-Flat, LO-FAT \cite{lofat2017} uses the same general principles and implements the attestation on a low-end embedded processor. Instead of an instrumentation-based approach with the requirement for some form of trusted execution environment, this approach enables attestation in hardware. This eliminates some drawbacks of C-Flat, such as the need for instrumentation and the performance overhead caused by it.

More recent works have further explored the \ac{cfa} concept both with \cite{litehax2018,atrium2017,lahel2020} and without \cite{recfa2021,tinycfa2021} hardware modifications.
Some are focusing on smaller embedded platforms \cite{tinycfa2021}. Tiny-CFA \cite{tinycfa2021} enables \ac{cfa} by utilizing APEX \cite{apex2020}, a technique to generate proofs of remote execution (PoX).
Other approaches are targeted at more complex architectures and programs \cite{scarr2019,recfa2021}. 
ScaRR \cite{scarr2019} introduces checkpoints via instrumentation and keeps a database of valid measurements that are compared to the measurements for the verification.
ReCFA \cite{recfa2021} presents another instrumented approach to enable \ac{cfa} for more complex architectures. It focuses on the additional challenges of more complex software, where the generated \acp{cfg} tend to be bigger, which makes it impractical to compute the hash values of all valid paths in advance. Instead, the entire control flow path that was observed is transmitted in compressed form.

The closest related work is presented by GuaranTEE \cite{guarantee2023}, where the \ac{cfa} is also based on two enclaves, where one enclave is responsible for attesting the control flow of the other. However, they are using the classic instrumentation approach to generate control flow traces directly via trampolines similar to C-Flat. These trampolines are then transmitting control flow events to their VerifyTEE, that is responsible for checking the validity of the control flow. 
Our goal was to cleanly separate the trace generation from the potentially compromised code, therefore, we decided against depending on components of the tracee for the trace generation.
Another difference is their creation of the \ac{cfg}, where they are repeatedly executing the traced code in the pre-processing phase in order to collect all valid edges. 
This has the downside that, if not all viable edges are reported, the attestation could potentially fail for valid control flows. 
Lastly, by locking any communication with the host application, our design \textbf{prevents} any form of leakage in the event of detected control flow violations.

Another related line of work is investigating the use of \acp{hpc} for \ac{rop} detection \cite{hdrop2014,sigdrop2016,ropsentry2017}. The detection of these approaches generally monitors certain \ac{hpc} to detect patterns related to \ac{rop} attacks, e.g. return mispredictions. 
Like our work, these works present concepts for the detection of \ac{rop} attacks. However, these approaches are only designed to detect patterns in measured \ac{hpc} values, which gives an indication that a \ac{rop} attack is happening, whereas our approach takes into account the \ac{cfg} of a program in combination with accurate \ac{hpc} predictions.
In addition to enabling the detection of very minute control flow violations (see \Cref{sec:reliability-metrics}) our approach is designed to detect all control flow violations, not just those introduced by \ac{rop} or \ac{jop} attacks.

\section{Conclusion} \label{sec:conclusion}
We have presented a concept to enable \ac{cfa} for \acp{tee}, which is based on two enclaves running in parallel. It is able to stop control flow attacks before these can leak sensitive data. 
Furthermore, we have demonstrated how the \acp{hpc} architecture can be utilized for a trace generation that is fully independent of the traced enclave. 
We pair this trace generation with the ability to accurately predict the effect that code sequences have on the \ac{hpc} trace to introduce \hpcfa.
Our proof-of-concept implementation of \hpcfa demonstrates the soundness of the design and is able to attest the control flow of valid programs without raising any alarms. 
Our evaluation uncovers a trade-off between the reliability and the performance overhead of our \ac{cfa} approach. 
With appropriate instrumentation of the traced code, our last experiments show how near perfect reliability can be achieved. 

\section*{Acknowledgements}
This work was supported by the German BMFTR project SASVI.

\bibliographystyle{IEEEtran}
\bibliography{references} 

\end{document}